\documentclass[10pt,aps,prb,twocolumn,amsmath,amssymb,superscriptaddress]{revtex4-1}
\pdfoutput=1
\usepackage{amsmath}
\usepackage{graphicx}
\graphicspath{ {./} }
\usepackage[caption = false]{subfig}
\usepackage{threeparttable}
\usepackage{color}

\def\beq{\begin{eqnarray}}
\def\eeq{\end{eqnarray}}
\def\V{\mathcal{V}}
\def\C{\mathcal{C}}
\def\Evar{E_{\mathrm{var}}}
\def\Etot{E_{\mathrm{tot}}}
\newcommand{\mtnote}[1]{\textsuperscript{\TPTtagStyle{#1}}}

\begin{document}

\title{Excited States of Methylene, Polyenes, and Ozone\\from Heat-bath Configuration Interaction}

\author{Alan D. Chien}
\affiliation{Department of Chemistry, University of Michigan, Ann Arbor,MI 48109,USA}

\author{Adam A. Holmes}
\affiliation{Department of Chemistry and Biochemistry, University of Colorado Boulder, Boulder, CO 80302, USA}
\affiliation{Laboratory of Atomic and Solid State Physics, Cornell University, Ithaca, NY 14853, USA}

\author{Matthew Otten}
\affiliation{Laboratory of Atomic and Solid State Physics, Cornell University, Ithaca, NY 14853, USA}

\author{C. J. Umrigar}
\affiliation{Laboratory of Atomic and Solid State Physics, Cornell University, Ithaca, NY 14853, USA}

\author{Sandeep Sharma}
\affiliation{Department of Chemistry and Biochemistry, University of Colorado Boulder, Boulder, CO 80302, USA}

\author{Paul M. Zimmerman}
\affiliation{Department of Chemistry, University of Michigan, Ann Arbor,MI 48109, USA}

\begin{abstract}
The electronically excited states of methylene (CH$_2$), ethylene (C$_2$H$_4$), butadiene (C$_4$H$_6$), 
hexatriene (C$_6$H$_8$), and ozone (O$_3$) have long proven challenging due to their complex mixtures of 
static and dynamic correlations. Semistochastic heat-bath configuration interaction (SHCI), which efficiently and 
systematically approaches the full configuration interaction (FCI) limit, is used to provide close approximations 
to the FCI energies in these systems. This article presents the largest FCI-level calculation to date -- on 
hexatriene using a polarized double-zeta basis (ANO-L-pVDZ), which gives rise to a Hilbert space containing 
more than $10^{38}$ determinants. These calculations give vertical excitation energies of 5.58 and 5.59 eV respectively for the
$2^1{\rm A}_{\rm g}$ and $1^1{\rm B}_{\rm u}$ states, showing that they are nearly degenerate.
The same excitation energies in butadiene/ANO-L-pVDZ were found to be 6.58 and 6.45 eV.
In addition to these benchmarks, our calculations strongly support the presence of a
previously hypothesized ring-minimum species of ozone that lies 1.3 eV higher than the open-ring
minimum energy structure and is separated from it by a barrier of 1.11 eV.
\end{abstract}

\maketitle

\section{Introduction}

The exponential increase in Hamiltonian dimension with increasing system size means that exact,
Born-Oppenheimer electronic energies are not easily achievable for polyatomic molecular
systems.~\cite{Knowles1984,Olsen1988,Olsen1990,Olsen1996,Rossi1999,Dutta2003,Gan2006}
Recent years have seen impressive progress in methods that produce FCI-quality energies at greatly
reduced cost, such as density matrix renormalization group
(DMRG),~\cite{White1993,White1999,Chan2002,Chan2011,Olivares-Amaya2015} FCI quantum Monte Carlo
(FCIQMC),~\cite{Booth2009,Cleland2010,PetHolChaNigUmr-PRL-12,Booth2013} and incremental FCI
(iFCI).~\cite{Zimmerman2017,Zimmerman2017a,Zimmerman2017b}

Another avenue for obtaining FCI-quality energies has also recently become available due to the revival
of the selected configuration interaction plus perturbation theory (SCI+PT) algorithms. SCI+PT consists of two steps.
In the first step, the most important determinants of the wavefunction of interest are identified
iteratively and the Hamiltonian in this subspace ($\mathcal{V}$) is diagonalized to obtain the
approximate variational energies and wavefunctions.
In the second step, the perturbative step of SCI+PT attempts to correct these energies and wavefunctions.
The first such SCI+PT method was called Configuration Interaction
by Perturbatively Selecting Iteratively (CIPSI), which established the basic steps of the SCI+PT
algorithms.~\cite{Huron1973,Buenker1975,Buenker1978,Evangelisti1983,Harrison1991,BenAmor2011}
Since then, many variations of CIPSI have been developed over the years,~\cite{BuePey-TCA-74,langlet1976excited,oliveros1978ci,cimiraglia1985second,cimiraglia1987recent,Knowles89,Har-JCP-91,povill1992treating,SteWenWilWil-CPL-94,GarCasCabMal-CPL-95,WenSteWil-IJQC-96,Neese-JCP-03,NakEha-JCP-05,AbrShe-CPL-05,BytRue-CP-09,Rot-PRC-09,Eva-JCP-14,Kno-MP-15,SchEva-JCP-16,LiuHof-JCTC-16,zhang2016deterministic,scemama2016quantum,garniron2017hybrid,giner2017jeziorski,schriber2017adaptive}
all of which try to improve upon the CIPSI algorithm.
However, a common drawback of all these variants is that, to construct the selected space
$\mathcal{V}$ iteratively, the algorithm has to loop over all the determinants connected to any
determinant in previous iteration of $\mathcal{V}$.
This becomes prohibitively expensive as the size of space $\mathcal{V}$ increases to several
million determinants.
Some of us have recently proposed the Heat-bath CI (HCI)~\cite{Holmes2016} algorithm that eliminates
this expensive step by changing the selection criterion such that it enables an algorithm that
loops over only those determinants that will be included in the selected space $\mathcal{V}$,
a small fraction of all the possible connected determinants.
This results in orders of magnitude speed up over other variants of SCI+PT for this step of the algorithm.
HCI was further improved by semistochastic evaluation of the pertubative energy in semistochastic HCI (SHCI),~\cite{Sharma2017}
which eliminated the need to store a long list of perturbative determinants in memory. SHCI's potential has
been demonstrated in previous works, where it was shown to efficiently treat CI spaces orders of magnitude larger than is
possible with conventional CI algorithms.~\cite{Sharma2017,HolUmrSha-JCP-17,Smith2017,Mussard17}

\begin{figure}
\centering
\includegraphics[scale = 0.8]{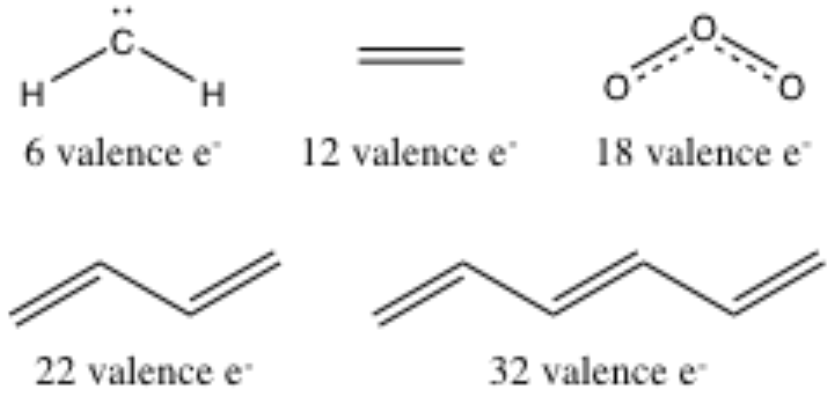}
\caption{The molecules, methylene, ethylene, ozone, butadiene and hexatriene, investigated with SHCI.}
\label{fig:mols}
\end{figure}

In this paper, highly accurate benchmarks for electronically excited states of the polyatomics in
Figure~\ref{fig:mols} are computed using SHCI. Methylene is presented as the first test case, due to its small
size yet challenging electronic structure.~\cite{Sherrill1998} Additionally, ozone is examined to answer a long-standing
question regarding the existence of a theorized meta-stable ozone species.~\cite{Hay1972,Theis2016}
Finally, SHCI is applied to the first few polyenes $-$ ethylene, butadiene,
and hexatriene $-$ which have long been studied for their role as prototypical organic conducting polymers.
The state ordering of the low-lying valence excited states, $2^1{\rm A}_{\rm g}$ and $1^1{\rm B}_{\rm u}$, in
butadiene and hexatriene have been especially challenging due to the significant numbers of highly correlated
electrons and the near-degenerate nature of the valence
states.~\cite{Tavan1987,Watts1996,Starcke2006,Li1999,Nakayama1998,Schreiber2008,Mazur2009,Schmidt2012,Piecuch2015}
Herein, extrapolated SHCI state energies will provide high accuracy $2^1{\rm A}_{\rm g}$/$1^1{\rm B}_{\rm u}$
state orderings in butadiene and hexatriene.

This article is organized as follows.
In Section II, the excited-state SHCI algorithm is reviewed, and the path to convergence to the FCI
limit is discussed. In Section III, results on the smaller methylene and ethylene systems are used to validate
SHCI against current benchmark values and establish convergence with respect to FCI. These observations
are then used to estimate FCI-quality energies for the larger ozone, butadiene, and hexatriene molecules.
Section IV provides conclusions and an outlook on the SHCI method.

\section{Methods}
\subsection{Semistochastic Heat-Bath Configuration Interaction}
\label{sec:SHCI}
As HCI and semistochastic perturbation theory have been described in
detail,~\cite{Holmes2016,Sharma2017,HolUmrSha-JCP-17} only a brief overview will be given here.
The HCI algorithm can be divided into variational and perturbative stages, each of which selects determinants
through threshold values, $\epsilon_{\rm V}$ and $\epsilon_{\rm PT}$, respectively.
The current variational space of determinants is denoted by $\V$ and the space of all determinants connected by single
or double excitations to $\V$, but not in $\V$, is denoted by $\C$.

The variational stage iteratively adds determinants to $\V$ by
\begin{enumerate}
\item Adding all determinants $a$ connected to determinants in the current $\V$ that pass the importance criterion $\max_i \left|H_{ai}
\max_n\left(\left|c_i^n\right|\right)\right| > \epsilon_{\rm V}$, where $c_i^n$ is the coefficient of determinant $i$ in state $n$.
\item Constructing the Hamiltonian and solving for the roots of interest, in the basis of all determinants in the newly expanded $\mathcal{V}$.
\item Repeat 1-2 until convergence.
\end{enumerate}

Convergence of the variational wavefunction for a given $\epsilon_{\rm V}$ is signified by the addition of a small number
of new determinants or small changes in the variational energy ($E_{\rm var}$). The second-order Epstein-Nesbet perturbative
energy correction ($\Delta E_2$) is added to $E_{\rm var}$ to obtain the total HCI energy ($E_{tot}$). This
correction is
\beq
\Delta E_2 = \sum_{a\in\mathcal{C}} \frac{\left(\sum_{i\in\mathcal{V}}^{(\epsilon_{\rm PT})} H_{ai} c_i\right)^2}{E_0-H_{aa}},
\eeq
where $a$ runs over determinants in $\mathcal{C}$, and $i$ over determinants in $\mathcal{V}$. Similar to
the variational stage, the perturbation only considers the determinants connected to the final $\V$ space
that have an importance measure greater than a parameter $\epsilon_{\rm PT}$, which is typically orders of
magnitude smaller than $\epsilon_{\rm V}$.  In both the variational and the perturbative stages, the fact that
the number of \emph{distinct} values of the double-excitation matrix elements scales only as $N_{\rm orb}^4$
is used to avoid ever looking at the unimportant determinants.
Nevertheless, storing the full space of determinants used in the perturbative
correction becomes a memory bottleneck for larger systems.

SHCI sidesteps this memory bottleneck using a semistochastic second-order perturbation correction.~\cite{Sharma2017}
In this procedure, the perturbative correction is split into deterministic and stochastic contributions.
A larger $\epsilon_{\rm PT}^{\rm d}$, automatically determined to correspond to a determinant space of
manageable size depending on available computer memory, is first used to obtain a deterministic energy correction. The remaining correlation is
then calculated stochastically by taking the difference of the second-order corrections evaluated with
$\epsilon_{\rm PT}$ and $\epsilon_{\rm PT}^{\rm d}$. Samples are taken until the statistical error falls below a
specified threshold.

\subsection{Converging SHCI Energies to the FCI Limit}
The target accuracy for total or relative energies are typically be chosen to be 1 mHa or 1.6 mHa
(1 kcal/mol, representing chemical accuracy), though for the smaller systems it is easy to achieve much higher accuracy.
In SHCI, the error in the variational energy can be straightforwardly estimated by the magnitude of the perturbative correction.

In methylene and ethylene, SHCI can provide such highly converged variational energies.
In larger systems, however, converging the variational energy would require prohibitively large variational spaces.
Instead, we fit the variational energy or the total energy, $\Etot = \Evar + \Delta E_2$, to $\Evar-\Etot$ using a quadratic function and
use the fitted function to extrapolate to the no perturbative correction ($\Evar-\Etot=0$) limit.
The fit coefficients for variational and total energies are the same, except that the coefficients of the linear terms differ by one,
so the two energies extrapolate to precisely the same value.
There is not a well-defined method for estimating the extrapolation error, but a reasonable choice is one fifth of the
difference between the calculated energy with the smallest value of $\epsilon_{\rm V}$ and the extrapolated energy.
In many cases the fitted function is very nearly linear (see e.g. Figs.~\ref{fig:ethFitE}, \ref{fig:butaExtrap} and \ref{fig:hexaExtrap}).
Furthermore, even when the fitted functions are not close to linear, the functions for different states are often close to parallel,
making the estimates of the energy differences particularly accurate.

\subsection{Computational Details}
SHCI is implemented in Fortran90, parallelized using MPI, and makes use of spatial symmetry, and
time-reversal symmetry when the number of up- and down-spin electrons is equal.~\cite{Sharma2017}
The variational iterations are terminated when
when the number of new determinants added is less than 0.001\%
of the current variational space or when the change in variational energy is less than $1\cdot10^{-5}$ Ha. For all
calculations, $\epsilon_{\rm PT}$ is set to $1\cdot10^{-7}$ Ha, which provides converged perturbative
corrections.~\cite{Holmes2016,Sharma2017} $\epsilon_{\rm V}$ is made as small as possible on our hardware,
obtaining either small $\Delta E_2$ or enough data points to reliably extrapolate to $\Delta E_2=0$.
The threshold for statistical error of the stochastic perturbative correction is generally set to $5\cdot10^{-5}$ Ha,
although the larger hexatriene/ANO-L-pVDZ computations use $1\cdot10^{-4}$ Ha.

For the smaller systems (methylene and ethylene) achieving convergence is relatively easy, allowing the use of Hartree-Fock 
and HCI natural orbitals (obtained with $\epsilon_{\rm V}=3\cdot 10^{-5}$ Ha), for methylene and ethylene respectively, 
to construct the molecular orbital integrals.
For the larger systems (ozone, butadiene, hexatriene), the convergence was improved by using orbitals that minimize
the HCI variational energy~\cite{Smith2017} for $\epsilon_{\rm V}=2\cdot10^{-4}$.
Possibly, yet better convergence could be obtained by using orbitals that make the total energy stationary~\cite{Smith2017}.

Basis sets used are aug-cc-pVQZ~\cite{Dunning1989,Kendall1992} for methylene, ANO-L-pVTZ~\cite{Widmark1990}
for ethylene, ANO-L-pVDZ~\cite{Widmark1990} for butadiene and hexatriene, and cc-pVTZ~\cite{Dunning1989}
for ozone. Geometries for methylene are FCI/TZVP quality taken from Sherrill et al.~\cite{Sherrill1998} and ozone
geometries are CASSCF(18,12)/cc-pVQZ quality, taken from Theis et al.~\cite{Theis2016} For the polyenes, all geometries are
of MP2/cc-pVQZ quality, with ethylene and hexatriene geometries the same as in Zimmerman~\cite{Zimmerman2017a} 
and butadiene the same as in Alavi et al.~\cite{Daday2012} and Chan et al.~\cite{Olivares-Amaya2015} 
All calculations utilize the frozen-core approximation. For comparisons to coupled cluster theories, the same
geometries and basis sets are used with the Q-Chem 4.0~\cite{Krylov2013} CR-EOM-CC(2,3)D~\cite{Woch2006}
implementation.~\cite{Manohar2008}

\section{Results and Discussion}

\subsection{Methylene}
Methylene is a prototypical test case for advanced electronic structure methods, being small enough to be amenable to
canonical FCI benchmarks, yet still requiring accurate treatment of dynamic and static correlations for correct excitation
energies.~\cite{Schaefer1986,Bauschlicher1986,Sherrill1997,Sherrill1998,Zimmerman2009,Slipchenko2002,Shao2003,Chien2017}
The four lowest lying states of methylene vary in spin and spatial symmetry: $1^3{\rm B}_{\rm 1}$, $1^1{\rm A}_{\rm 1}$,
$1^1{\rm B}_{\rm 1}$, and $2^1{\rm A}_{\rm 1}$. With only six valence electrons to correlate, SHCI can handily obtain
FCI-quality energies even with the large aug-cc-pVQZ basis (Table~\ref{table:methE}), obtaining perturbative corrections
less than 0.01 mHa with $\epsilon_{\rm V}$ = $10^{-5}$ Ha.

Table~\ref{table:methE} shows the most accurate SHCI adiabatic energy gaps calculated with $\epsilon_{\rm V}$ = $10^{-5}$ Ha,
which differ from experiment by about 0.01 eV. Comparing canonical FCI in the TZ2P basis with SHCI in the larger
aug-cc-pVQZ basis shows differences of up to 0.158 eV,~\cite{Sherrill1998} demonstrating that large basis sets are
necessary to fully describe correlation in methylene. This was first demonstrated using diffusion Monte Carlo (DMC)
results,~\cite{Zimmerman2009} which are much less sensitive to basis sets and agree with SHCI to within about 0.02 eV.
CR-EOMCC(2,3)D relative energies are generally within 1.6 mHa (0.044 eV) of the benchmark SHCI values, indicating
that high-level multi-reference coupled cluster calculations are able to correlate six electrons sufficiently to obtain FCI-quality energy gaps.

\begin{table*}
\caption{Methylene/aug-cc-pVQZ total (Ha) and relative (eV) energies}
\centering
\begin{threeparttable}
\begin{tabular}{c|c|c|c|c|c}
\hline
State & SHCI\mtnote{a} & CR-EOMCC (2,3)D\mtnote{c} & FCI\mtnote{b} \\
      & aug-cc-pVQZ    & aug-cc-pVQZ & TZ2P          \\
\hline
$1^3$B$_1$ & -39.08849(1) & -39.08817 & -39.06674\\
$1^1$A$_1$ & -39.07404(1) & -39.07303 & -39.04898\\
$1^1$B$_1$ & -39.03711(1) & -39.03450 & -39.01006\\
$2^1$A$_1$ & -38.99603(1) & -38.99457 & -38.96847\\
\hline
\hline
Gap & SHCI\mtnote{a} & CR-EOMCC (2,3)D\mtnote{c} & FCI\mtnote{b} & DMC\mtnote{d} & Exp\\
    & aug-cc-pVQZ    & aug-cc-pVQZ               & TZ2P          &               &    \\
\hline
$1^1$A$_1-1^3$B$_1$ & 0.393 & 0.412 & 0.483 & 0.406 & 0.400\mtnote{e}\\
$1^1$B$_1-1^3$B$_1$ & 1.398 & 1.460 & 1.542 & 1.416 & 1.411\mtnote{f}\\
$2^1$A$_1-1^3$B$_1$ & 2.516 & 2.547 & 2.674 & 2.524 & $-$\\
\hline
\end{tabular}
\begin{tablenotes}
  \item[a] Using $\epsilon_{\rm V}$ = $10^{-5}$ Ha
  \item[b] FCI/TZ2P results from reference~\citenum{Sherrill1998}
  \item[c] This work
  \item[d] Diffusion Monte Carlo results from reference~\citenum{Zimmerman2009}
  \item[e] References~\citenum{Sherrill1998},~\citenum{Jensen1988}
  \item[f] References~\citenum{Sherrill1998},~\citenum{Alijah1990}
\end{tablenotes}
\end{threeparttable}
\label{table:methE}
\end{table*}

\subsection{Ethylene}
Ethylene is another prototypical benchmark system for electronic excitations, including an especially
challenging $1^1{\rm B}_{\rm u}$ state. Although the $1^1{\rm B}_{\rm u}$ state is qualitatively well
described by a $\pi$-$\pi$* excitation, a quantitative description requires a thorough accounting of
dynamic correlation between $\sigma$ and $\pi$ electrons.~\cite{Davidson1996,Muller1999,Angeli2010}
Here, SHCI is applied to the low-lying valence states of ethylene: $1^1{\rm A}_{\rm g}$, $1^1{\rm B}_{\rm 1u}$
and $1^3{\rm B}_{\rm 1u}$, in the ANO-L-pVTZ basis.

\begin{figure}
\begin{center}
\includegraphics[width=0.5\textwidth]{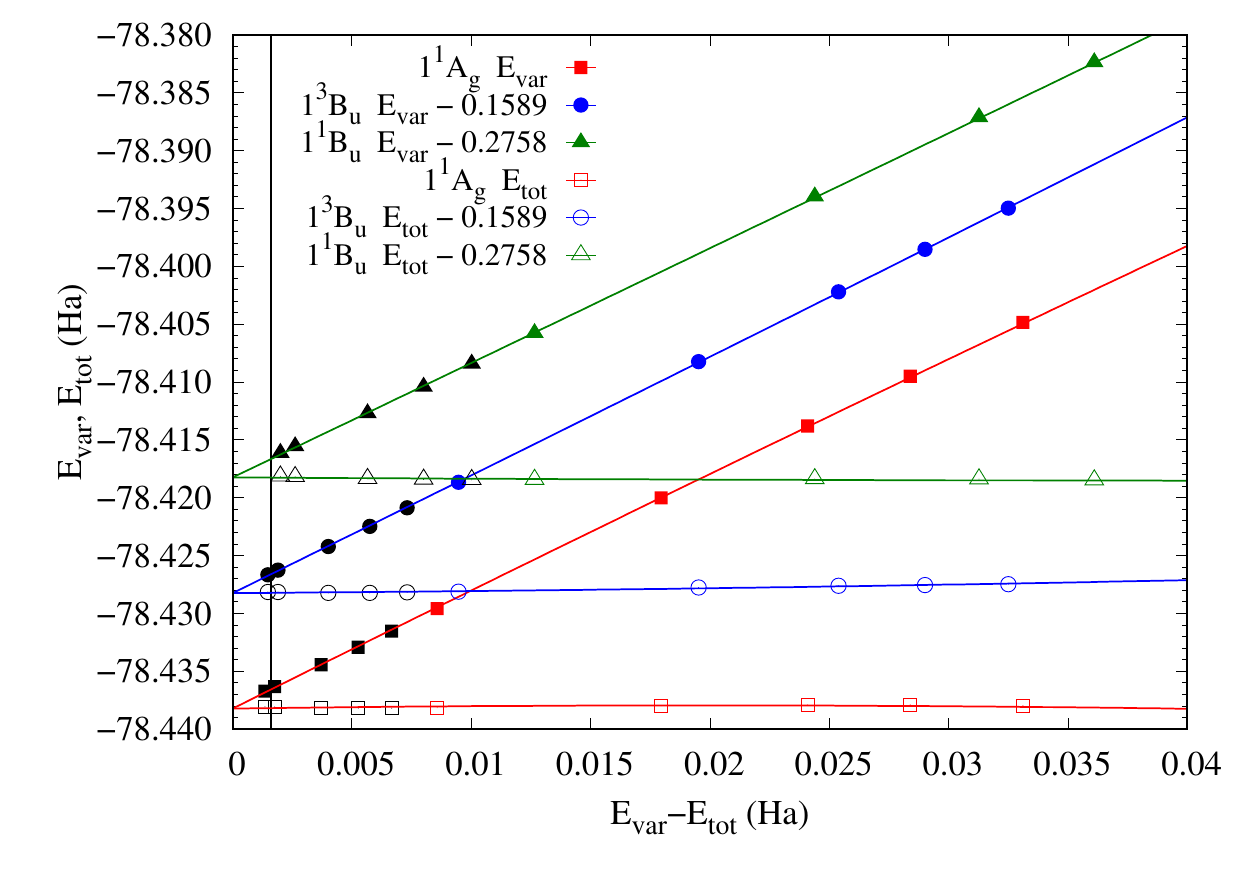}
\end{center}
\caption{Fit of variational and total energies of ethylene using all data points.
The black line is placed at 1.6 mHa. States are separated from one another by 0.01 Ha for clarity.
The tightest SHCI calculation used $\epsilon_{\rm V}$ = $7\cdot10^{-6}$ Ha.
The extrapolated energies obtained using all data points are compared to those obtained by omitting the black points
in Table~\ref{table:ethFitE}.}
\label{fig:ethFitE}
\end{figure}

Fully correlating ethylene's twelve valence electrons is a considerably more difficult task than correlating
methylene's six. This is reflected in the fact that SHCI perturbative corrections start to fall below 1.6 mHa
only at $\epsilon_{\rm V}$ = $7\cdot10^{-6}$ Ha (Figure~\ref{fig:ethFitE}).
These results suggest that polyatomics with up to twelve valence electrons and triple-zeta basis sets are amenable to
treatment at the FCI level using just the variational component of SHCI.
Table~\ref{table:ethE} compares SHCI total and relative energies with previous FCIQMC~\cite{Daday2012} and
iFCI~\cite{Zimmerman2017a} results.
SHCI total energies are only about 1 mHa lower than FCIQMC and the $1^1$B$_{\rm 1u}-1^1$A$_{\rm g}$ excitation energy
is in even better agreement.
The $1^3$B$_{\rm 1u}-1^1$A$_{\rm g}$ energy obtained from iFCI is also in reasonably good agreement, thought it is
obtained using a different triple-zeta basis.
On the other hand, Table~\ref{table:ethE} also indicates that coupled cluster methods
must include more than triples excitations in order to obtain FCI-quality relative energies, as
CR-EOMCC(2,3)D results show errors considerably greater than 1.6 mHa (0.044 eV) with respect to the SHCI benchmark values.
The SHCI relative energies support the notion that the vertical excitations cannot be quantitatively compared to
the experimental band maxima in ethylene.~\cite{Daday2012,Zimmerman2017b}

\begin{table*}
\caption{Ethylene/ANO-L-pVTZ total (Ha) and relative (eV) energies}
\begin{threeparttable}
\begin{tabular}{c|c|c|c|c|c}
\hline
State & SHCI\mtnote{a} & CR-EOMCC(2,3)D\mtnote{b} & FCIQMC\mtnote{c}\\
      & ANO-L-pVTZ     & ANO-L-pVTZ & ANO-L-pVTZ      \\
\hline
$1^1$A$_{\rm g}$ & -78.4381(1) & -78.43698 & -78.4370(2)\\
$1^1$B$_{\rm 1u}$ & -78.1424(1) & -78.13375 & -78.1407(3)\\
$1^3$B$_{\rm 1u}$ & -78.2693(1) & -78.26205 & - \\
\hline
\hline
Gap & SHCI\mtnote{a} & CR-EOMCC(2,3)D\mtnote{b} & FCIQMC\mtnote{c} & iFCI\mtnote{d} & Exp\\
    & ANO-L-pVTZ     & ANO-L-pVTZ     & ANO-L-pVTZ       & cc-pVTZ        &    \\
\hline
$1^1$B$_{\rm 1u}-1^1$A$_{\rm g}$ & 8.05 & 8.25 & 8.06 & - & 7.66\mtnote{e}\\
$1^3$B$_{\rm 1u}-1^1$A$_{\rm g}$ & 4.59 & 4.76 & - & 4.64 & 4.3-4.6\mtnote{f}\\
\hline
\end{tabular}
\begin{tablenotes}
  \item[a] $E_{\rm tot}$ with $\epsilon_{\rm V}$ = $7\cdot10^{-6}$ Ha
  \item[b] This work
  \item[c] FCIQMC/ANO-L-pVTZ results from reference~\citenum{Daday2012}
  \item[d] iFCI/cc-pVTZ results from reference~\citenum{Zimmerman2017a}
  \item[e] Experimental band maximum from reference~\citenum{Mulliken1977}
  \item[f] Experimental band maxima from references~\citenum{Moore1970,Moore1972,VanVeen1976}
\end{tablenotes}
\end{threeparttable}
\label{table:ethE}
\end{table*}

Ethylene is the largest system tested for which the perturbative correction is less than 1.6 mHa. This requires using
$\epsilon_{\rm V}$=$7\cdot10^{-6}$ Ha and $10^8$ determinants in the variational space, which is
near the limit of what can be reasonably stored on contemporary hardware.

As mentioned in Methods: Converging SHCI Energies to the FCI Limit, in larger systems we fit 
$\Etot$ or $\Evar$ to $\Evar-\Etot$ using a quadratic function and
use the fitted function to extrapolate to the no perturbative correction limit, thereby obtaining accurate
energies even when the variational energies are not converged.~\cite{HolUmrSha-JCP-17}
The fits of $\Etot$ and $\Evar$ are shown in Fig.~\ref{fig:ethFitE}.  The $\Etot$ is nearly flat.
To estimate the error of the extrapolation, we performed an additional fit omitting the black points in Fig.~\ref{fig:ethFitE}.
The extrapolated values obtained from the two fits are shown in Table~\ref{table:ethFitE}.

\begin{table}
\caption{Comparison of extrapolated  ethylene/ANO-L-pVTZ energies obtained using all the points plotted
in Fig.~\ref{fig:ethFitE} with those obtained from omitting the black points.}
\centering
\begin{tabular}{c|c|c}
\hline
State & $E_{\rm extrap}$ (Ha) & $E_{\rm extrap}$ (Ha) \\
      & all points            & omit black points     \\
\hline
\hline
$1^1$A$_{\rm g}$  & -78.4382 & -78.4385 \\
$1^1$B$_{\rm 1u}$ & -78.1424 & -78.1430 \\
$1^3$B$_{\rm 1u}$ & -78.2693 & -78.2697 \\
\hline
\end{tabular}
\label{table:ethFitE}
\end{table}

\subsection{Ozone}
\begin{figure}
\centering
\begin{center}
\includegraphics[scale=0.55]{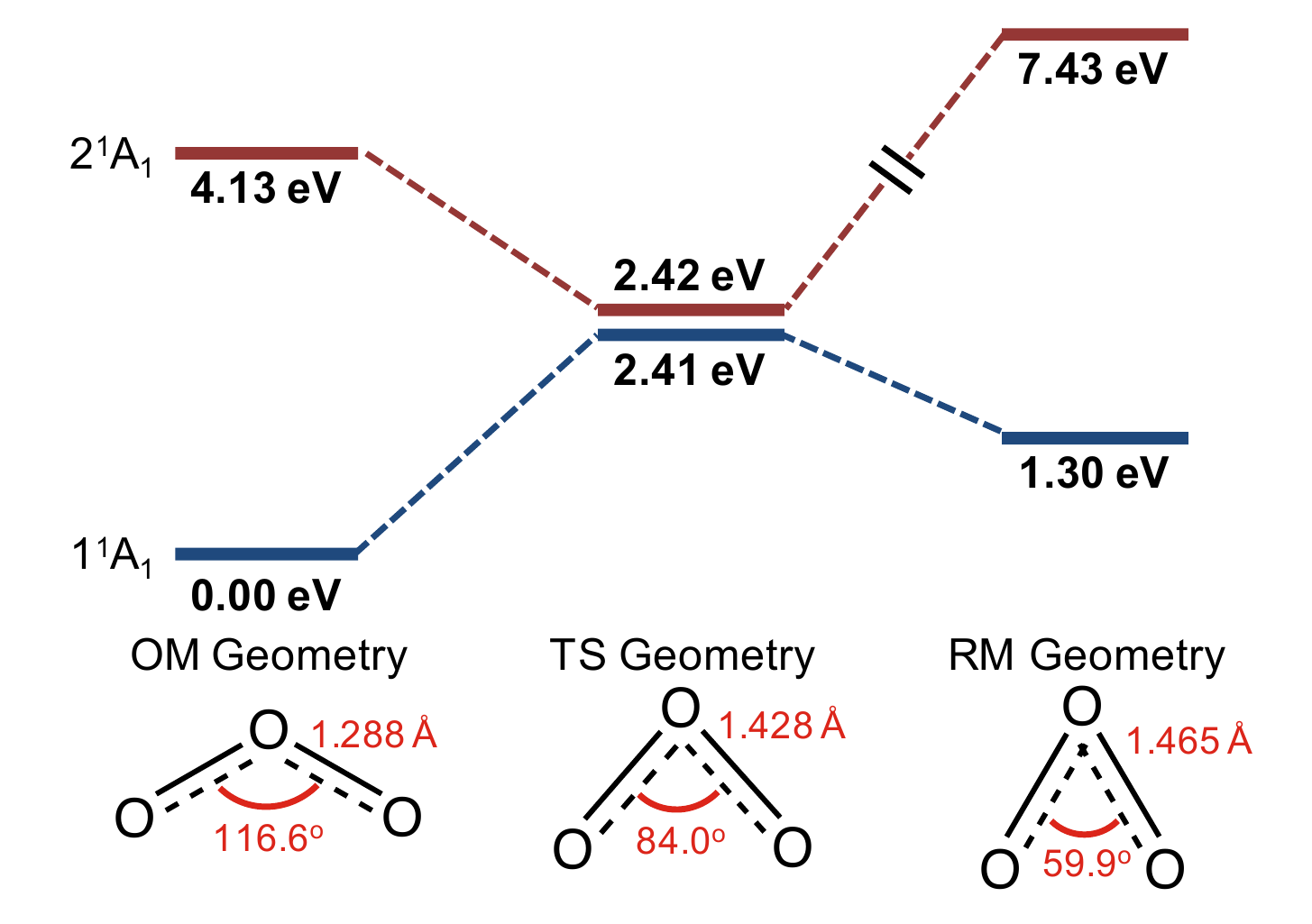}
\caption{Ozone potential energy surface}
\label{fig:ozone}
\end{center}
\end{figure}

Ozone's potential energy surfaces have held great interest due to its role in atmospheric
chemistry.~\cite{Andersen2013} An interesting feature predicted by computational studies is the
existence of a metastable ring geometry on the ground state surface.~\cite{Hay1972} A lack of
experimental evidence for such a species has fueled multiple studies of the pathway leading to
the ring species over the years.~\cite{Lee1990,Qu2005,Xantheas1990,Xantheas1991,Atchity1997,Atchity1997a}
The most recent such study by Ruedenberg et al. utilizes multi-reference CI with up to quadruple
excitations,~\cite{Theis2016} expending considerable effort on selecting and justifying an active space.
To provide an accurate picture at critical points along the theorized pathway with even treatment
of all valence electrons, SHCI is applied to ozone's $2^1{\rm A}_{\rm 1}$-$1^1{\rm A}_{\rm 1}$ gap
with the cc-pVTZ basis at the three geometries of interest shown in Figure~\ref{fig:ozone}: the equilibrium
geometry (termed the open ring minimum (OM)), the hypothetical ring minimum (RM), and the
transition state (TS) between these two.

As anticipated, sub-mHa perturbative corrections cannot be readily obtained for ozone in the cc-pVTZ basis.
$\Delta E_2$ for the best available SHCI calculations, at $\epsilon_{\rm V}$=$4\cdot10^{-5}$ Ha, range from
15-28 mHa for the various geometries and states under consideration. The accuracy of ozone's extrapolated
$2^1{\rm A}_{\rm 1}$-$1^1{\rm A}_{\rm 1}$ gaps can easily be corroborated for the OM and TS geometries,
as the gaps over a broad range of $\epsilon_{\rm V}$'s (Table~\ref{table:ozConv}) vary by less than 1 mHa,
so the extrapolated values should be even more accurate.
The RM geometry's gap is not as easily corroborated, as these vary over a 2.1 mHa range at reasonably tight $\epsilon_{\rm V}$'s. 
Therefore, a conservative view would be to take the extrapolated gap as slightly less than chemically accurate.

In Table~\ref{table:ozGapE}, the SHCI energy gaps are compared to Ruedenberg et al's MRCI results.~\cite{Theis2016}
SHCI results mostly resemble the MRCI estimates, except for the RM geometry, where the gaps differ by more than 1 eV.
The SHCI results, however, are sufficiently converged to allow valuable insights to be made into the
meta-stable nature of the RM species. Along the $1^1{\rm A}_{\rm 1}$ potential surface, the RM and TS geometries
lie 1.30 eV and 2.41 eV, respectively, above the OM geometry. These values suggest that electronic excitations in
ozone are likely required to reach RM, but that the RM species should be relatively stable with a 1.11 eV barrier
hindering return to the OM geometry. Thus, SHCI indicates that a RM species may well exist, and that experimental
investigations should be able to observe it if a plausible isomerization pathway can be accessed.

\begin{table}
\caption{Evolution of ozone $2^1{\rm A}_{\rm 1}$ - $1^1{\rm A}_{\rm 1}$ gaps (Ha).
The RM energy at $\epsilon_{\rm V}=2\cdot10^{-4}$ is omitted from the fit.}
\begin{tabular}{c|c|c|c}
\hline
$\epsilon_{\rm V}$ & OM & TS & RM\\
\hline
\hline
$2\cdot10^{-4}$  & 0.1520 & 0.0009 & 0.2897\\
$1\cdot10^{-4}$ & 0.1522 & 0.0008 & 0.2314\\
$5\cdot10^{-5}$ & 0.1521 & 0.0005 & 0.2298\\
$4\cdot10^{-5}$ & 0.1521 & 0.0006 & 0.2293\\
\hline
Extrapolated & 0.1519 & 0.0003 & 0.2254\\
\hline
\end{tabular}
\label{table:ozConv}
\end{table}

\begin{table}
\caption{Ozone $2^1{\rm A}_{\rm 1}$ - $1^1{\rm A}_{\rm 1}$ gaps (eV)}
\begin{threeparttable}
\begin{tabular}{c|c|c}
\hline
Geometry & Extrapolated SHCI & MRCI (SDTQ)\mtnote{a}\\
\hline
\hline
OM & 4.13 & 3.54-4.63\\
TS & 0.01 & 0.05-0.16\\
RM & 6.13 & 7.35-8.44\\
\hline
\end{tabular}
\begin{tablenotes}
  \item[a] Reference \citenum{Theis2016}
 \end{tablenotes}
\end{threeparttable}
\label{table:ozGapE}
\end{table}

\subsection{Shorter Polyenes: Butadiene and Hexatriene}
Butadiene and hexatriene are part of the polyene series, long studied for their role as prototypical organic
conducting polymers. In particular, the spacing of the low-lying valence excited states has proven especially
challenging for electronic structure methods.~\cite{Tavan1987,Watts1996,Starcke2006,Li1999,Nakayama1998,Mazur2009,Schmidt2012,Schreiber2008,Piecuch2015}
Butadiene and hexatriene are of special interest because their $1^1{\rm B}_{\rm 1u}$ and $2^1{\rm A}_{\rm g}$
states are nearly degenerate, resulting in conflicting reports of state ordering at lower levels of theory. In the
ANO-L-pVDZ basis, butadiene and hexatriene's FCI spaces of $10^{26}$ and $10^{38}$ determinants,
respectively, are too large for the routine application of FCI-level methods, although limited FCIQMC~\cite{Daday2012}
and DMRG~\cite{Olivares-Amaya2015} studies as well as SHCI ground state calculations~\cite{HolUmrSha-JCP-17}
have been performed on butadiene. Herein, SHCI is applied to the $1^1{\rm A}_{\rm g}$, $1^1{\rm B}_{\rm 1u}$,
$1^3{\rm B}_{\rm 1u}$, and $2^1{\rm A}_{\rm g}$ states to provide accurate benchmarks and state orderings.

\subsubsection{Butadiene}
\begin{figure}
\begin{center}
\includegraphics[width=0.5\textwidth]{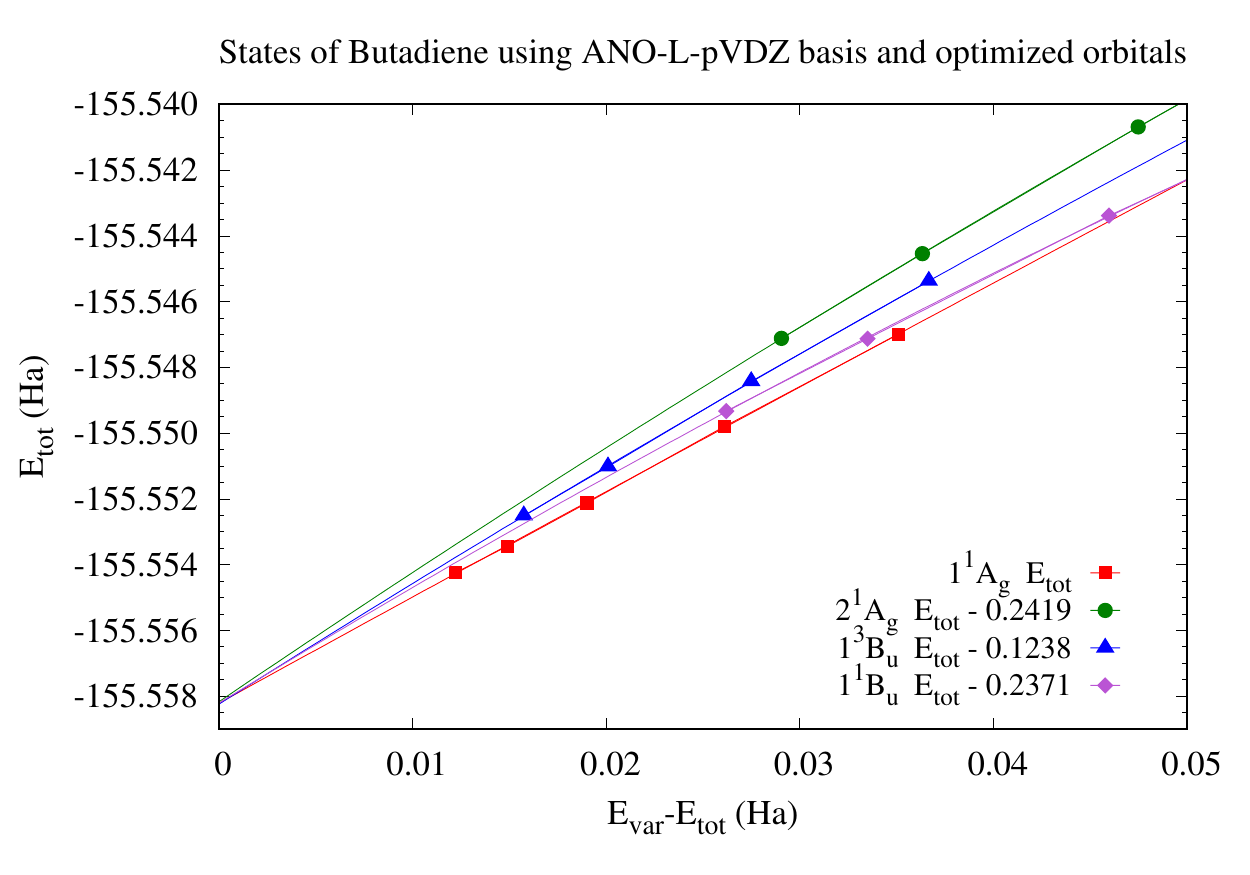}
\caption{Extrapolation of butadiene SHCI energies. States shifted to extrapolate to the same energy. The tightest SHCI calculation used $\epsilon_{\rm V}$ = $3\cdot10^{-5}$ Ha.}
\label{fig:butaExtrap}
\end{center}
\end{figure}

Similar to ozone, extrapolation is used to obtain FCI energy estimates for butadiene in the ANO-L-pVDZ
basis, as the tightest SHCI calculations at $\epsilon_{\rm V}$ = $3\cdot10^{-5}$
Ha had perturbative corrections ranging from 12-29 mHa (Figure~\ref{fig:butaExtrap}).
Besides using orbitals that minimize the SHCI variational energy, for molecules with more than a few atoms a further
improvement in the energy convergence can be obtained by localizing the orbitals.
Butadiene has C$_{\rm 2h}$ symmetry, but the localized orbitals transform as the irreducible representations of the C$_{\rm s}$
subgroup of C$_{\rm 2h}$.  Both the A$_{\rm g}$ and the B$_{\rm u}$ irreducible representations of C$_{\rm 2h}$ transform as the A'' representation
of C$_{\rm s}$.  Hence calculating the three singlet states, $1^1{\rm A}_{\rm g}$, $2^1{\rm A}_{\rm g}$ and $1^1{\rm B}_{\rm 1u}$
would require calculating three states simultaneously if localized orbitals are used, and further we would not know
if the $2^1{\rm A}_{\rm g}$ or the $1^1{\rm B}_{\rm 1u}$ is lower in energy.  Consequently, we calculated only
the $1^1{\rm A}_{\rm g}$ and $1^3{\rm B}_{\rm 1u}$ orbitals using localized orbitals and computed the
$2^1{\rm A}_{\rm g}$ and $1^1{\rm B}_{\rm 1u}$ states with extended orbitals.

Table~\ref{table:butaGapE} shows that using the same geometry as in prior FCIQMC,~\cite{Daday2012}
DMRG,~\cite{Olivares-Amaya2015} and iFCI~\cite{Zimmerman2017b} calculations leads to a SHCI $1^1{\rm A}_{\rm g}$ energy 
that is 0.4 and 0.9 mHa below the extrapolated DMRG~\cite{extrap_Butadiene_DMRG} and iFCI energies respectively.
Although FCIQMC has yielded very accurate energies for many systems, in the case of butadiene
all three methods (SHCI, iFCI, and DMRG) are in agreement that the FCIQMC energy for the $1^1{\rm A}_{\rm g}$ is
8-9 mHa too high.  This may be either because of FCIQMC initiator bias or because of an underestimate of
the FCIQMC statistical error because of the very long auto-correlation times encountered for large systems.
A similar conclusion can be reached 
for the FCIQMC $1^1{\rm B}_{\rm 1u}$ calculation, as the SHCI energy falls below it by a large amount, 12 mHa.

Turning to relative energies, we see that SHCI is in close agreement with all prior FCI-level theoretical calculations.
Both the $1^3{\rm B}_{\rm u}$-$1^1{\rm A}_{\rm g}$  and $1^1{\rm B}_{\rm u}$-$1^1{\rm A}_{\rm g}$ gaps are 0.08 eV
away from the iFCI and FCIQMC values respectively.
Although the $2^1{\rm A}_{\rm g}$-$1^1{\rm A}_{\rm g}$ gap does not currently have FCI-level benchmarks, the agreement
of SHCI's other relative energies with existing benchmarks supports the accuracy of the extrapolated SHCI
value for this gap, which is 6.58 eV. This places the $2^1{\rm A}_{\rm g}$ state above $1^1{\rm B}_{\rm u}$
in butadiene by 0.13 eV.  This small gap is consistent with recent theoretical~\cite{Komainda2017} and
experimental~\cite{Fuss2001} investigations demonstrating ultrafast population transfer from $1^1{\rm B}_{\rm u}$
to $2^1{\rm A}_{\rm g}$, which implies close proximity of the two states.
As with ethylene, relative energies only qualitatively agree with experiment, supporting prior indications that
experimental band maxima of butadiene do not correspond to the vertical excitation energy.~\cite{Watson2012}

\begin{table}
\caption{Butadiene total (Ha) and relative energies (eV).
Extrapolation errors may range from a few tenths of a mHa for $1^1{\rm A}_{\rm g}$ to a couple of mHa for $2^1{\rm A}_{\rm g}$.}
\label{table:butaGapE}
\begin{threeparttable}
\begin{tabular}{c|c|c|c|c}
\hline
State & Extrapolated SHCI & FCIQMC\mtnote{a} & DMRG\mtnote{b}\\
\hline
$1^1$A$_{\rm g}$ & -155.5582(1) & -155.5491(4) & -155.5578\\
$1^3$B$_{\rm u}$ & -155.4344(1) & - & - \\
$1^1$B$_{\rm u}$ & -155.3211(1) & -155.3092(6) & - \\
$2^1$A$_{\rm g}$ & -155.3163(1) & - & -\\
\hline
\hline
Gap & Extrapolated SHCI & FCIQMC\mtnote{a} & iFCI\mtnote{c} & Exp \\
\hline
$2^1{\rm A}_{\rm g}$-$1^1{\rm A}_{\rm g}$ & 6.58 & - & - & -  \\
$1^1{\rm B}_{\rm u}$-$1^1{\rm A}_{\rm g}$  & 6.45 & 6.53 & - & 5.92\mtnote{d}\\
$1^3{\rm B}_{\rm u}$-$1^1{\rm A}_{\rm g}$ & 3.37 & - & 3.45 & 3.22\mtnote{e}\\
\hline
\end{tabular}
\begin{tablenotes}
  \item[a] FCIQMC/ANO-L-pVDZ results from reference \citenum{Daday2012}
  \item[b] DMRG/ANO-L-pVDZ extrapolated energy~\cite{extrap_Butadiene_DMRG} using data from reference \citenum{Olivares-Amaya2015}
  \item[c] iFCI 6-31G* results from reference \citenum{Zimmerman2017a}
  \item[d] Experimental band maxima from references \citenum{Mosher1973,McDiarmid1976,Doering1980}
  \item[e] Experimental band maxima from reference \citenum{Mosher1973a}
\end{tablenotes}
\end{threeparttable}
\end{table}

\subsubsection{Hexatriene}
\begin{figure}
\begin{center}
\includegraphics[width=0.5\textwidth]{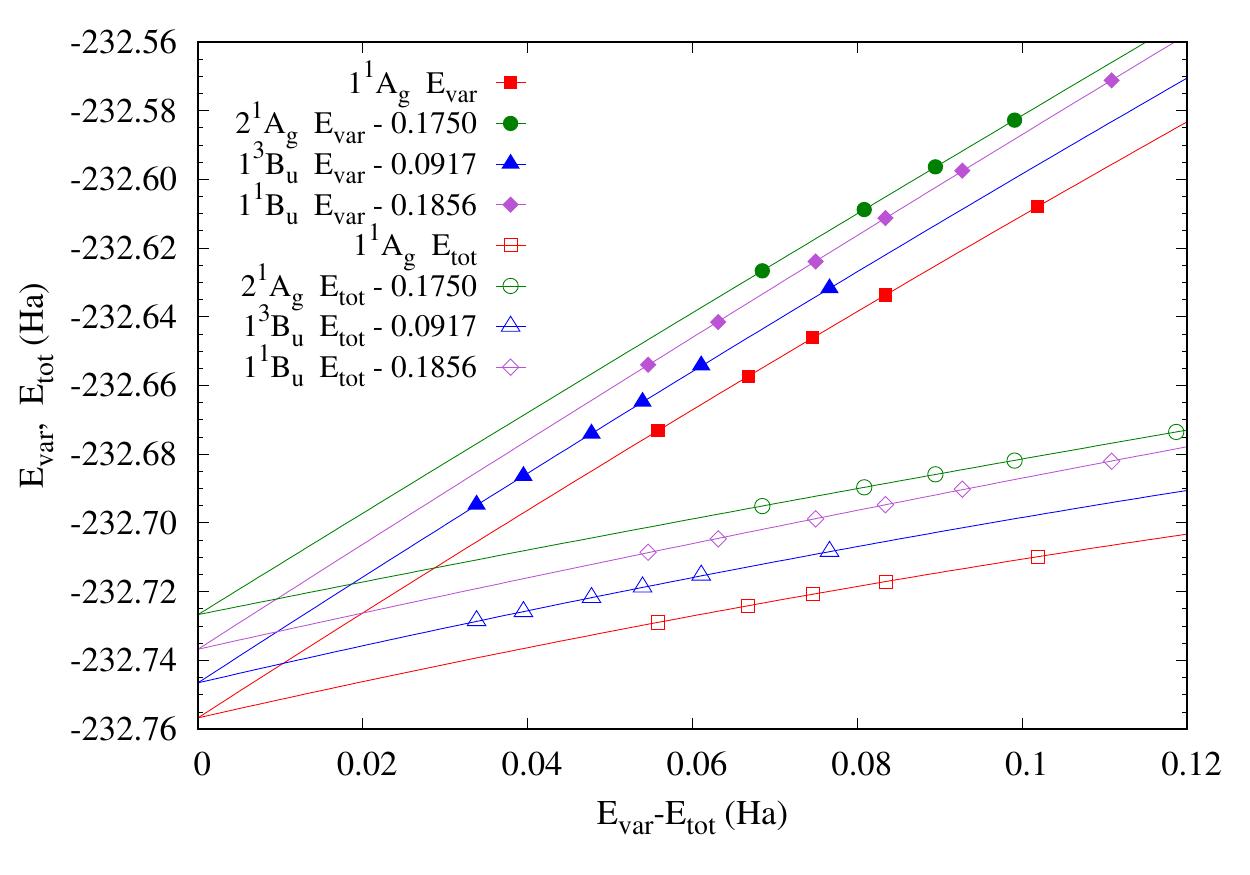}
\caption{Extrapolation of hexatriene SHCI energies. States are separated from each other by 0.01 Ha for clarity.
The tightest SHCI calculation used $\epsilon_{\rm V}$ = $2\cdot10^{-5}$ Ha.}
\label{fig:hexaExtrap}
\end{center}
\end{figure}

Hexatriene is at the current frontier of FCI-level computations, with a demanding FCI space of $10^{38}$
determinants in the ANO-L-pVDZ basis. Only one other algorithm, iFCI,~\cite{Zimmerman2017a} has approached
FCI energies for such a large polyatomic - and then only for its singlet-triplet gap.
Here we compute the energies of the lowest three singlet states and the lowest triplet state,
using values of $\epsilon_{\rm V}$ as small as $2\cdot10^{-5}$ Ha (Figure~\ref{fig:hexaExtrap}), 
which results in as many as $9\cdot10^{7}$ variational determinants.

The extrapolated hexatriene energies are reported in Table~\ref{table:hexGapE}.
With the same geometry, SHCI produces a $1^1{\rm A}_{\rm g}$ total energy 4 mHa below iFCI.
This difference is within the extrapolation uncertainty of SHCI for this system.
Prior investigations of hexatriene photo dynamics~\cite{Hayden1995,Cyr1996,Ohta1998} place $1^1{\rm B}_{\rm u}$ 
close in energy to $2^1{\rm A}_{\rm g}$. At the vertical excitation geometry, SHCI places $2^1{\rm A}_{\rm g}$ below
$1^1{\rm B}_{\rm u}$ with a small gap of only 0.01 eV.
The triplet-singlet $1^3{\rm B}_{\rm u}$-$1^1{\rm A}_{\rm g}$ gaps computed by SHCI and
iFCI (using the slightly smaller 6-31G* basis) differ by only 0.04 eV.
As in the case of butadiene, the SHCI $1^1{\rm B}_{\rm u}$-$1^1{\rm A}_{\rm g}$ gap differs significantly from
experiment,~\cite{Komainda2016} indicating that experimental band maxima do not correspond to vertical
excitation energies in hexatriene.

\begin{table}
\centering
\caption{Hexatriene total (Ha) and relative energies (eV)}
\begin{threeparttable}
\begin{tabular}{c|c|c|c|c}
\hline
State & Extrapolated SHCI & & iFCI\\
\hline
$1^1$A$_{\rm g}$ & -232.7567(1) & & -232.7527\mtnote{c}\\
$1^3$B$_{\rm u}$ & -232.6548(1) & & - \\
$1^1$B$_{\rm u}$ & -232.5511(1) & & - \\
$2^1$A$_{\rm g}$ & -232.5517(1) & & - \\
\hline
\hline
Gap & Extrapolated SHCI & CC & iFCI & Exp\\
\hline
$2^1{\rm A}_{\rm g}$-$1^1{\rm A}_{\rm g}$ & 5.58 &5.72\mtnote{a} & - & 5.21\mtnote{e}\\
$1^1{\rm B}_{\rm u}$-$1^1{\rm A}_{\rm g}$ & 5.59 &5.30\mtnote{a} & - & 4.95\mtnote{e},5.13\mtnote{e}\\
$1^3{\rm B}_{\rm u}$-$1^1{\rm A}_{\rm g}$ & 2.77 &2.80\mtnote{b} & 2.81\mtnote{d} & 2.61\mtnote{f}\\
\hline
\end{tabular}
\begin{tablenotes}
  \item[a] CR-EOMCC(2,3)D/TZVP from reference \citenum{Piecuch2015}
  \item[b] CCSD(T)/6-31G* from reference \citenum{Zimmerman2017a}
  \item[c] iFCI/ANO-L-pVDZ result from reference \citenum{Zimmerman2017b}
  \item[d] iFCI/6-31G* result from reference \citenum{Zimmerman2017a}
  \item[e] Raman scattering results from reference \citenum{Fujii1985}
  \item[f] Electron impact band maximum from reference \citenum{Flicker1977}
\end{tablenotes}
\end{threeparttable}
\label{table:hexGapE}
\end{table}

\section{Conclusion}
SHCI represents an important step forward for SCI methods, providing FCI-quality energies in the largest molecular
systems to date. SHCI easily correlates systems of 12 electrons in a triple-zeta basis and can reach FCI-level
energies in larger systems through extrapolation. In this paper, CI spaces of $10^7-10^8$ determinants were used
to effectively handle FCI spaces of $10^{26}$ and $10^{38}$ determinants.

This work has provided new benchmarks and insights for the valence states of some commonly investigated
molecular systems. Specifically, high-quality SHCI energetics for ozone give strong evidence that the
theorized RM structure has a significant barrier to relaxation, and thus should be observable by experiment.
Investigation of butadiene and hexatriene lead to the highest level $2^1{\rm A}_{\rm g}$/$1^1{\rm B}_{\rm u}$
state-orderings in these systems to date, placing $2^1{\rm A}_{\rm g}$ above $1^1{\rm B}_{\rm u}$ in butadiene,
and minutely below $1^1{\rm B}_{\rm u}$ in hexatriene. In short, SHCI has shown itself to
be an efficient means of obtaining FCI-level energetics, and we look forward to it providing physical insight into other
chemically interesting systems with up to dozens of electrons.

\begin{acknowledgements}
ADC and PMZ would like to thank David Braun for computational support and the University of Michigan for financial support.
SS acknowledges the startup package from the University of Colorado.
AAH, MO and CJU were supported in part by NSF grant ACI-1534965.
Some of the computations were performed on the Bridges computer at the Pittsburgh Supercomputing Center,
supported by NSF award number ACI-1445606.
\end{acknowledgements}

\bibliographystyle{jchemphys}
\bibliography{SHCI_paper,hci}

\end{document}